\documentclass[]{article}
\usepackage{lscape}
\usepackage{pdflscape}
\usepackage{graphicx}
\usepackage{multirow}
\usepackage{dcolumn}
\usepackage[numbers]{natbib}
\usepackage{amsmath}
\usepackage{amssymb}
\usepackage{listings}
\usepackage[T1]{fontenc}
\usepackage[utf8]{inputenc}
\usepackage{graphicx}
\usepackage{epstopdf}
\usepackage{color}
\usepackage{float}
\usepackage{tikz}
\usepackage{tcolorbox}
\usepackage{multirow}
\usepackage{hhline}
\usepackage{xcolor}
\usepackage{colortbl}
\usepackage{subcaption}
\usepackage{booktabs}
\usepackage{lscape}
\usepackage{subcaption}
\usepackage{pgfplots}
\usepackage{lineno}
\usepackage{listings}
\usepackage[utf8]{inputenc}
\usepackage[letterpaper,pdftex]{geometry}
\usepackage{longtable}
\usetikzlibrary{arrows,automata,fit,scopes,calc,matrix,positioning,decorations.pathmorphing,decorations.pathreplacing}
\pgfplotsset{width=9cm,compat=1.15}
\usepgfplotslibrary{statistics}
\usepackage[all]{xy}
\usepackage[affil-it]{authblk}
\usepackage[doublespacing]{setspace}
\usepackage{verbatim}
\usepackage{rotating}

\usepackage{hyperref}	
\hypersetup{colorlinks,linkcolor={blue},citecolor={blue!50!black},urlcolor={blue}}
\usepackage{geometry}

\author[1,2]{Federico Fioravanti \footnote{federico.fioravanti9@gmail.com}}
\author[1,3]{Fernando Delbianco \footnote{fernando.delbianco@uns.edu.ar}} 
\author[1,3]{Fernando Tohm\'e \footnote{ftohme@criba.edu.ar}}

\affil[1]{Instituto de Matem\'atica de Bah\'ia Blanca, CONICET - UNS, Bah\'{\i}a Blanca, Argentina}
\affil[2]{Departamento de Matem\'atica, Universidad Nacional del Sur, Bah\'ia Blanca, Argentina}
\affil[3]{Departamento de Econom\'ia, Universidad Nacional del Sur, Bah\'ia Blanca, Argentina}

\title{The Relative Importance of Ability, Luck and Motivation in Team Sports: a Bayesian Model of Performance in the English Rugby Premiership}
\date{}
\begin{document}
\maketitle
\begin{abstract}
	Results in contact sports like Rugby are mainly interpreted in terms of the ability and/or luck of teams. But this neglects the important role of the {\em motivation} of players, reflected in the effort exerted in the game. Here we present a Bayesian hierarchical model to infer the main features that explain score differences in rugby matches of the English Premiership Rugby 2020/2021 season. The main result is that, indeed, {\em effort} (seen as a ratio between the number of tries and the scoring kick attempts) is highly relevant to explain outcomes in those matches.
\end{abstract}
\section{Introduction}
The development of mathematical models of sports faces many obstacles. Assessing the potential impact of unobservable variables and establishing the right relations among the observable ones are the main sources of hardships for this task. Even so, the study of team sports data has become increasingly popular in the last years. Several models have been proposed for the estimation of the parameters (characteristics) that may lead to successful results for a team, ranging  from machine learning methods to predict outcomes \cite{vstrumbelj2012simulating,asif2016play,baboota2019predictive}, fuzzy set representations \cite{hassanniakalager2020conditional}, statistical models \cite{dyte2000ratings,goddard2005regression,boshnakov2017bivariate} and Bayesian models \cite{baio2010bayesian,constantinou2012pi,wetzels2016bayesian,santos2019bayesian}.\\

One of the main issues in the study of sports is to disentangle the relative relevance of the possible determinants of outcomes. While {\em ability} and {\em luck} constitute, at least for both  the press and the fan base, the main explanatory factors of the degree of success in competitions, the motivation of players is usually invoked only to explain epic outcomes or catastrophic failures. One possible reason for the neglect of motivation is that, unlike ability and luck, it is hard to assess. In this paper we define a particular notion of {\em effort} in Rugby games as a proxy for motivation and develop a Bayesian model of the final scores of teams in the English Rugby Premiership 2020/2021. These outcomes will be explained by several variables, among which we distinguish the ability of the teams and the effort exerted by them. We also include as explanatory variables other possible sources of psychological stimuli, as to capture a pure motivation to win, separated from those other factors.\\

One of the main advantages of using Bayesian techniques to model sports are that beliefs or expert information can be incorporated as priors, to obtain posterior distributions of the parameters of interest, easily updated when new data becomes available, dealing more effectively with small data sets. In our case, we propose a Bayesian hierarchical model to explain score differences on a rugby match, i.e, the difference between home team points and away team ones. The main parameters of the model are the ability of teams, the effort exerted by them and the advantage (or disadvantage) of home teams.\\

There are many papers that use Bayesian methods to model the score of a rugby game. Stefani \cite{stefani2009predicting} finds that the past performance is better predictor of score difference than of the score total, and suggest that teams should focus strategy on score differences (to win or draw) rather than in score total. Pledger and Morton \cite{pledger2011modelling} use Bayesian methods to model the 2004 Super Rugby competition and explore how home advantage impacts the outcomes. Finally, Fry et al. \cite{fry2021variance} propose a Variance Gamma model where analytical results are obtained for match outcomes, total scores and the awarding of bonus points. The main difference between these works and ours, is that their primary goal is to predict outcomes, while ours is to explain them.\\

The \textit{ability} of a team can be conceived as its ``raw material''. The skills of its players, the expertise of its coaches and its human resources in general (medical staff, managers, etc.) constitute the team's basic assets. Their value can vary during a season due to injuries, temporary loss of skilled players called to play for the national team, players leaving the team, etc. In this model we assume that the capabilities of teams do not change much from a season to the next. Accordingly, the ability of a team at the start of the season is assumed to be at a bounded distance from the performance in the previous season.\\

\textit{Luck} in games and sports has been largely studied, from philosophical perspectives \cite{simon2007deserving,morris2015moral} to statistical ones \cite{denrell2012top,pluchino2018talent}. Maubossin \cite{mauboussin2012success} define that games that are high in luck are the ones that are highly unpredictable, it is not able to achieve great advantages through repetition and the `reversion to the mean' effect in performance is high. Elias et al. \cite{elias2012characteristics} and Gilbert and Wells \cite{gilbert2019ludometrics} define many types of luck, that we will later introduce. In this model, following the line from the later authors, we consider that luck is when the unexplained part differs markedly from the mean value in the noise distribution. That is, a series of unobserved variables have a huge impact on the outcome. \\

\textit{Effort}, in turn, can be conceived as the cost of performing at the same level over time and staying steadily engaged on a determinate task \cite{herlambang2021modeling}. Different measures of effort can be defined. In the context of decision making, effort can be the total number of elementary information processing operations involved \cite{payne1995trading} or the use of cognitive resources required to complete a task \cite{russo1983strategies,johnson1985effort}. Another measure of effort (or lack of it) can be defined in terms of the extent of {\em anchoring} in a self-reported rating scales, that is, the tendency to select categories in close proximity to the rating category used for the immediately preceding item \cite{lyu2022psychometric}. From a sports science perspective, the effort exerted by a team can be seen as the sum of all the players loads, that Quarrie et al. \cite{quarrie2017managing} define as `the total stressors and demands applied to the players'. These loads can include the physical motions of the players, the preparation for future matches, the food intakes, the intensity of interpersonal relationships, etc. Our formal definition of effort is intended as a proxy for the amount of some of these loads. Although a commonly accepted definition of effort is lacking, according to Massin \cite{massin2017towards} effort is understood as the force exerted in order to reach a goal. Our definition of effort is indeed intended to capture this notion, in the understanding that one of the main goals (and the hardest to achieve) in a rugby game is to score tries. We consider that deviating from the goal of scoring tries, that is, exerting forces in order to reach a different goal, implies a reduction of the effort.\\

In order to define a rough measure of the effort exerted by a rugby team, we follow the lead of Lenten and Winchester \cite{lenten2015secondary}, Butler et al. \cite{butler2020bonus} and Fioravanti et al. \cite{fioravanti2021effort}. These works analyze the effort exerted by rugby teams under the idiosyncratic incentives induced in this game. Besides gaining points for winning or drawing in a game, teams may earn ``bonus'' points depending on the number of times they score tries on a game. Accordingly, any appropriate effort measure should also be defined taking into account the number of tries. In our model the effort is measured as the ratio between the number of tries scored and the sum of tries and scoring kicks attempts\footnote{By scoring kicks we refer to conversions, penalty and drop kicks; and by kick attempt we mean every time the team decided to do a scoring kick, no matter the outcome.}. In other words, following the same ideas of the literature discussed above, we intend to measure the effort of a team using observable variables such as tries and kicks. Attempts to increase the score with tries instead than with kicks can be seen, following the definition of \cite{massin2017towards}, as indicating that the team is exerting more effort. Our idea is to emphasize on the identification of {\em effort} with the result of an offensive spirit, according to which a team maximizes this effort by seeking to get more tries, no matter what the final score is. But if we simply identify effort with the number of tries we run into a problem since the difference in scores (the dependent variable in our model) is highly correlated with the difference in tries. Considering instead the {\em proportion} of tries we do not run into that problem. Despite this, our formulation is not uncontroversial. It is easy to conceive a situation in which a team gets a lower value of our effort index by scoring more tries and kicks than a team that just scores one try with only one kick attempt. Again, our effort index intends to capture that tries involve an attacking, positive mindset, while penalties are a defensive, risk averse route to winning.\footnote{We are thankful to the Editor and two anonymous referees for pointing out that our definition of effort seems to be a strong proxy for  attack mindedness or risk seeking behavior.} We understand that this proxy of effort has its limitations, since it tends to disregard the defensive skills of the teams. Still, we can justify this choice by our goal of remaining close to the literature while keeping the model simple with minimal information requirements. It is worth to mention that asking teams to concentrate their efforts on scoring more tries is very intuitive albeit somewhat detrimental to the sport, as the uncertainty in results becomes highly reduced \cite{scarf2019outcome}. Similar counter-intuitive ideas have been also discussed for soccer in \cite{fry2021managing}. \\

Several studies detected the relevance of \textit{home advantage}, i.e. the benefit over the away team of being the home team. Schwartz and Barsky \cite{schwartz1977home} suggested that crowds exert an invigorating motivational influence, encouraging the home side to perform well. Still, a full explanation of this phenomenon requires taking into account the familiarity with the field of the home team, the travel fatigue of the away team, the social pressure exerted by the local fans over the referees, among other factors. Many other researchers investigated this advantage from different points of view, such as the physiological \cite{neave2003testosterone}, the psychological \cite{agnew1994crowd,legaz2013home}, the economic one \cite{carmichael2005home,boudreaux2017natural,ponzo2018does} and even exploring the possibility that referees may be favorably biased towards home teams \cite{downward2007effects,page2010evidence}. Home advantage in Rugby Union and Rugby League has been studied and confirmed by Kerr and van Schaik \cite{kerr1995effects}, Jones et al. \cite{jones2007home}, Page and Page \cite{page2010alone}, García et al. \cite{garcia2013home} and even during the Covid-19 pandemic by Fioravanti et al. \cite{fioravanti2021home}. In our model, home advantage is explored depending on if there is public allowed to attend the game, and in what day it is played\footnote{The tournament under consideration took place during the Covid-19 pandemic. Several games were played without attendance.}. An extra parameter intends to capture the influence of factors other than public attendance inducing home advantage.\\

The plan of the paper is as follows. In Section 2 we present the data of the English Premiership Rugby Championship, played in 2020/2021. Section 3 presents a Bayesian hierarchical model of the variables that explain the difference of scores in that championship. Section 4 runs a statistical descriptive analysis of the matches of the Premiership Championship in the light of the variables defined in the Bayesian model. Section 5 presents the results of estimating our model with the data of the Rugby Union competition. Section 6 considers the outliers found in the previous section, treating them as the result of luck in those games. We assess the aspects that justify considering them as instances of luck. Finally, Section 7 concludes and discusses the opportunities for further research.\\

\section{Data}
The Premiership Rugby Championship is the top English professional Rugby Union competition. The 2020/2021 edition was played by 12 teams. The league season comprises 22 rounds of matches, with each club playing each other home and away. The top 4 teams classify to the playoffs. Four points are awarded for the winning team, two to each team in case of a draw, and zero points to the loser team. However, a bonus point is given to the losing team in case the score difference is less than eight points. Teams also receive a bonus point in case they score four or more tries. In a game, a try is worth five points, a conversion two points and both penalty and drop kicks are worth three points each\footnote{After a try, the scoring team has the chance to kick for a conversion, except when a penalty-try is awarded. In the latter case seven points are automatically awarded to the team.}. During this season, if a game was canceled due to Covid-19, two points were awarded to the team responsible, and four to the other, while the match result was deemed to be $0-0$. The 2020/2021 season was won by the Harlequins, who claimed their second title after ending in the fourth league position.\\

The total score, number of tries, converted tries, converted penalties, attempted penalties, converted drops, attempted drops and attendance at each of the 122 games of the 2020/21 Premiership season have been taken from the corresponding Wikipedia entry\footnote{Ten games were canceled because some players tested positive for COVID 19. We do not consider the playoff games as we assume that the incentives are not the same at the elimination stage as in the classification games.}. We generate the priors of our Bayesian model based on the final ranking from the 2019/20 season as follows. An attack and defense ranking is built using the number of points scored and received by each team: the team with most tries scored and less points received is ranked first in both rankings. These rankings are then normalized, and their corresponding means are computed.  \\

\section{Model}
We base our model on a previous work of Kharratzadeh \cite{kharratzadeh2017hierarchical} that models the difference in scores for the soccer English Premier League. The score difference in game $g$, is denoted as $y_g$, and is assumed to follow  a $t_{student}$ distribution,
$$y_g\sim t_\nu (a_{diff}(g)+eff_{diff}(g)+ha(g),\sigma_y),$$
where $a_{diff}(g)$ is the difference in the ability of the teams, $eff_{diff}(g)$ is the difference in the effort exerted and $ha(g)$ the home advantage at game $g$. We give it a $N(0.5,1)$ prior. In turn, we assign a prior $Gamma(9,0.5)$ to the distribution of degrees of freedom $\nu$. \footnote{Kharratzadeh \cite{kharratzadeh2017hierarchical} uses a prior of $Gamma(2,0.5)$ based on the work of Ju\'arez and Steel \cite{juarez2010model}, where the shape parameter $2$ corresponds to a more positive skewed distribution with a smaller average difference in goals in the Premier League in soccer. Here we use a higher shape parameter corresponding to the less skewed distribution with a larger mean difference in scores  of the Premiership Rubgy Championship.}\\

We model the difference in abilities as follows:
$$a_{diff}(g)=a_{hw(g),ht(g)}-a_{aw(g),at(g)}$$
where $a_{hw(g),ht(g)}$ is the ability of the home team in the week where the game $g$ is played (analogously for the away team). We assume that the ability may vary during the season. More precisely, we assume that the ability at a period $t$ is the ability at $t-1$ plus a term representing the factors that may affect the ability at $t$:

$$a_{hw(g),ht(g)}=a_{hw(g)-1,ht(g)}+\sigma \cdot \eta_{hw(g),ht(g)}\, \mbox{for} \, hw\geq 2$$
where $\sigma$ and $\eta$ have weak informative priors $N(0,0.1)$ and $N(0,0.5)$, respectively. The model is analogous for the away team. The abilities for the first week depend on the previous performance of the teams, again assuming that ability has some sort of ``inertia'':
$$a_{1,ht(g)}=\beta_{prev}\cdot prevperf (ht(g))+\eta_{1,ht(g)} $$
where $\beta_{prev}$ is given the weakly informative prior $N(0.5,1)$ and $prevperf(j)$ is the previous performance of team $j$. This value is obtained as follows: we build two rankings of tries scored and received during the last season, where a team is at the top of both rankings if it  has scored the most and received the least tries in the last season. Then, the two rankings are normalized and averaged.\\

The variable that captures the relative {\em motivations} of the teams in the game is the difference in efforts:
$$ eff_{diff}(g)=\beta_{effort}\cdot(effH(g)-effA(g))$$
where $\beta_{effort}$ has a $N(0.5,1)$ prior and the effort of the home team is given an observational approximation  (analogously for the away team):
$$effH(g)=\dfrac{number\, of\, home\, tries\, in\, game\, g}{number\, of\, home\, tries\, in\, game\, g + attempted\,home\,scoring\, kicks\, in\, game\, g}.$$

Notice that the variable {\em attempted home scoring kicks in game $g$} has, in turn, three components. The number of conversions allowed after scoring tries, the number of penalty kicks attempted and the number of drop kicks attempted by the home team.\\

Our intention with this definition is to capture the idea that scoring tries demands more effort than other means of scoring points, and motivated teams try to maximize this value.\\

\begin{figure}[h!]
	
\begin{tikzpicture}[node distance={25mm},main/.style = {draw, circle,minimum size=12mm}] 
	\node[main] (1) {$\beta_{prev}$};
	\node[main](2) [right of =1]{$\eta$};
	\node[main](3) [right of =2]{$\sigma$};
	\node[main] (4) [right of =3]{$\beta_{effort}$};
	\node[main](5) [right of =4]{$\beta_{home}$};
	\node[main](6) [right of =5]{$\beta_{atten}$};
	\node[main] (7) [below of =2]{$a_{diff}$};
	\node[main](8) [below of =4]{$eff_{diff}$};
	\node[main](9) [right of =8]{$ha$};
	\node[main](10)[below of =8]{$y_g$};
	\draw[->] (1) -- (7);
	\draw[->] (2) -- (7);
	\draw[->] (3) -- (7);
	\draw[->] (4) -- (8);
	\draw[->] (5) -- (9);
	\draw[->] (6) -- (9);
	\draw[->] (7) -- (10);
	\draw[->] (8) -- (10);
	\draw[->] (9) -- (10);
\end{tikzpicture}
\caption{Graph corresponding to Model II}
\label{fig:graf}
\end{figure}
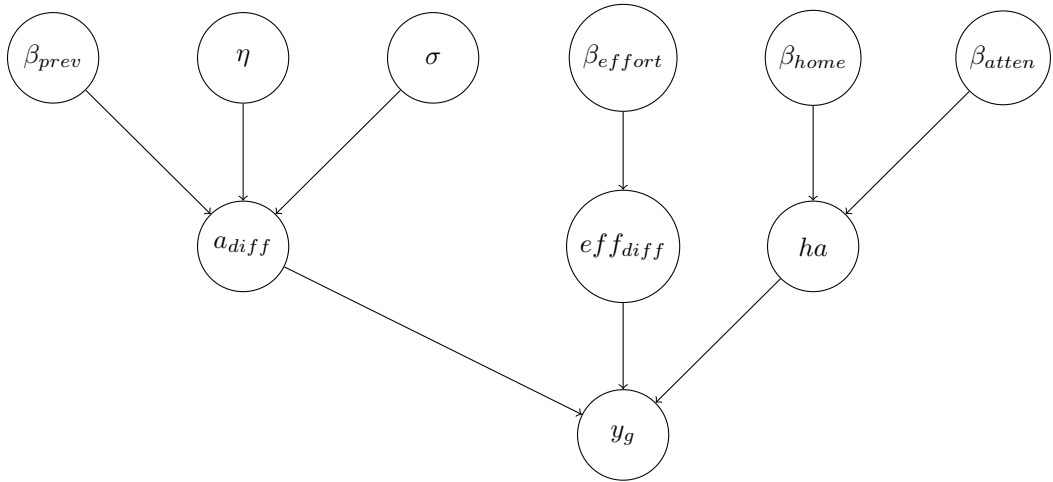

Finally, to capture home advantage, we consider both the attendance and non-attendance (such as the weather, long trips to play the game, etc.) effects .
$$ha(g)=\beta_{home}+\beta_{atten}\cdot atten(g),$$
where $\beta_{home}$ and $\beta_{atten}$ have $N(0.5,1)$ priors and $atten(g)$ is $0$ if no fans were allowed and $1$ otherwise. A graphical representation of the model is depicted in Figure \ref{fig:graf}.\\

To ensure robustness in our results we work with four different models. Model I does not include attendance as a variable of home advantage. Model II includes the attendance variable, while Model III, incorporates a day variable $day(g)$, with a $N(0.5,1)$ prior, which has value $1$ if the game was played on Saturday or Sunday and $0$ otherwise. This day variable allows to find out whether playing on a day in which almost all the fans can attend the game benefits either the home or the away team. Finally, Model IV includes a variation of $prevperf(j)$, where instead of the tries, we use total points scored and received by each team. There is no crucial difference between the four models. The reason for including different specifications is to evaluate whether the coefficients corresponding to the variables of interest, common to the four models, are sensitive to the inclusion of their specific variables.


\section{Descriptive Statistics}
On Table \ref{tab:home} and Table \ref{tab:away} we can see the descriptive statistics for home and away teams respectively where \textit{Score, T, C, P, D, AC, AP, AD} and {\it eff} indicate, respectively, total points, tries scored, conversions scored, penalty kicks scored, drop kicks scored, attempted conversions, attempted penalty kicks, attempted drop kicks and effort exerted.

\begin{table}[h!]
	\centering
	\caption{Home team}
	\label{tab:home}
	\begin{tabular}{rlllllllll}
		\hline
		&     ScoreH &       TH &       CH &       PH &       DH 	&      ACH &      APH &      ADH &effH\\ 
		\hline
		Min. & 3.00   & 0.00   & 0.00   & 0.00   & 0  & 0.00   & 0.00   & 0   & 0.00\\ 
		1st Qu & 17.00   & 2.00   & 1.00   &1.00   & 0  & 2.00   & 1.00   & 0   &0.28 \\ 
		Median & 23.00   & 3.00   & 2.00   & 2.00   & 0   & 3.00   & 2.00   & 0   &0.40 \\ 
		Mean & 26.22   & 3.248   & 2.407   & 1.708   & 0   & 3.186   & 2.027   & 0   &0.37 \\ 
		3rd Qu & 34.00   & 4.00   & 3.00   & 3.00   & 0   &  4.00   & 3.00   & 0   &0.46  \\ 
		Max. & 74.00   & 12.00   & 7.00   & 6.00   & 0   & 12.00   & 7.00   &0   & 0.55 \\ 
		\hline
	\end{tabular}
	\caption*{\textbf{Note}: H: Home team; T: tries; C: conversions; P: penalties; D: drops, A: prefix for attempted actions; eff: effort ratio.}
\end{table}

\begin{table}[h!]
	\centering
	\caption{Away team}
	\label{tab:away}
	\begin{tabular}{rlllllllll}
		\hline
		&     ScoreA &       TA &       CA &       PA &       DA&      ACA &      APA &      ADA& effA\\ 
		\hline
		Min &  3.00   & 0.00   & 0.00   & 0.00   & 0.00  &0.00   & 0.00   & 0.00&0.00 \\ 
		1st Qu. & 15.00   & 2.00   & 1.00   & 0.00   & 0.00  &0.00   & 0.00   & 0.00 &0.28 \\ 
		Median  & 22.00   & 3.00   & 2.00   & 1.00   & 0.00&3.00   & 1.00   & 0.00&0.37   \\ 
		Mean & 22.35   & 2.796   & 1.929   & 1.434   & 0.017 &2.717   & 1.655   & 0.026&0.36  \\ 
		3rd Qu. & 28.00   & 4.00   & 3.00   & 2.00   & 0.00 &4.00   & 3.00   & 0.00&0.5  \\ 
		Max. & 62.00   & 9.00   & 6.00   & 5.00   & 1.00 &8.00   & 5.00   & 2.00  &0.66\\ 
		\hline
	\end{tabular}
	\caption*{\textbf{Note}: A: away team; T: tries; C: conversions; P: penalties; D: drops, A: prefix for attempted actions; eff: effort ratio.}
\end{table}

\begin{figure}[hbt!]
	\centering
	\caption{Score Difference Histogram}
	\includegraphics[width=0.75\textwidth]{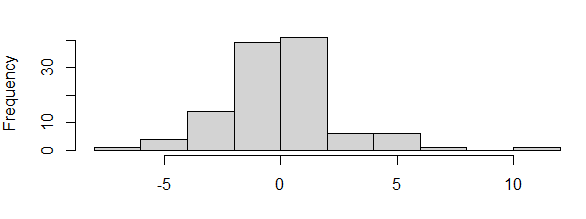}
	\label{fig:score_diff}
	
\end{figure}
\begin{figure}[hbt!]
	\centering
	\caption{Effort Histograms}		
	\begin{subfigure}[t]{0.47\textwidth}
		\centering
		\scriptsize
		\includegraphics[width=1\textwidth]{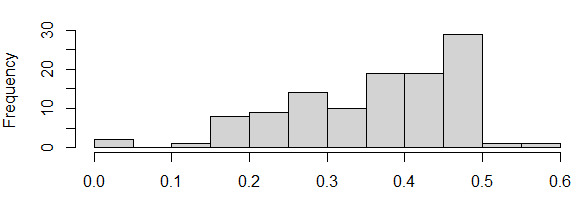}
		\caption{Effort Home Team. \label{fig:ratioH}}
	\end{subfigure}%
	~ 
	\begin{subfigure}[t]{0.47\textwidth}
		\centering
		\scriptsize
		\includegraphics[width=1\textwidth]{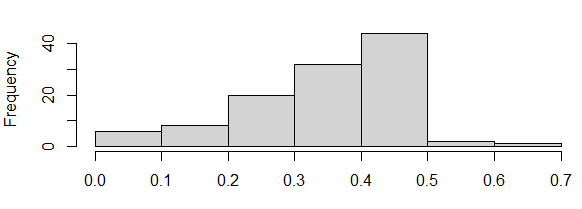}
		\caption{Effort Away Team. \label{fig:ratioA}}
	\end{subfigure}
\end{figure}

Figure \ref{fig:score_diff} shows that most of the score differences are around zero (even though there are very few draws obtained in the championship). This indicates that games usually end with little difference. Figures \ref{fig:ratioH} and \ref{fig:ratioA} depict the effort histograms. We can see great number of cases of $effort=\frac{1}{2}$. This is because teams almost always have the chance to go for a conversion after scoring a try, except when a penalty-try is awarded; and in many games the teams did not attempt to kick a penalty (maybe because they have no kickable penalties available).

\section{Results}

We estimate the model with the R package \textit{rstan} \cite{stanrstan}. We use 4 cores, each one to run 2500 iterations and 1500 warm-up ones. The Stan code of the estimated model can be found in the Appendix.\\
 
The results obtained appear to hold a degree of robustness: 
\begin{itemize}
\item The $Rhat$ (or Gelman-Rubin) statistic measures the discrepancies between the chains generated in simulations of Bayesian models. The further its value is from $1$, the worse. But we can see in all our results that $Rhat$ is very close to $1$.
\item $n_{eff}$ is an estimate of the effective sample (of parameters) size. A large value indicates a low degree of error in the expected value of the parameter. We can see that, indeed, this is the case for all the parameters of interest.
\end{itemize}

\begin{figure}[hbt!]
	\centering
	\caption{Histograms}		
	\begin{subfigure}[t]{0.47\textwidth}
		\centering
		\scriptsize
		\includegraphics[width=1\textwidth]{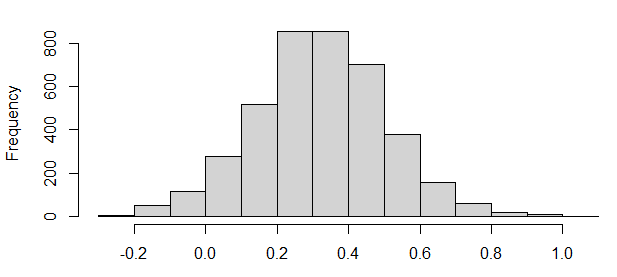}
		\caption{Beta Home. \label{fig:b_home}}
	\end{subfigure}%
	~ 
	\begin{subfigure}[t]{0.47\textwidth}
		\centering
		\scriptsize
		\includegraphics[width=1\textwidth]{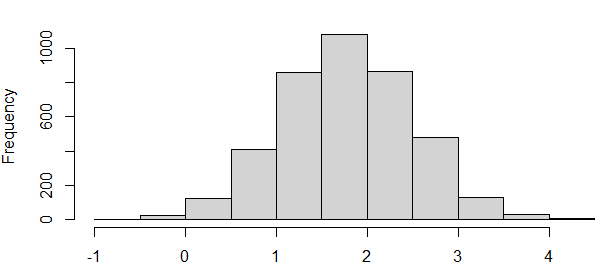}
		\caption{Beta Previous. \label{fig:b_prev}}
	\end{subfigure}
	\label{hist}
\end{figure}

\begin{figure}[hbt!]
	\centering
	\caption{Histograms (Cont.)}		
	\begin{subfigure}[t]{0.47\textwidth}
		\centering
		\scriptsize
		\includegraphics[width=1\textwidth]{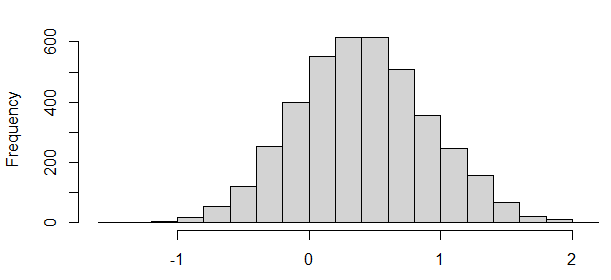}
		\caption{Beta Attendance. \label{fig:b_att}}
	\end{subfigure}%
	~ 
	\begin{subfigure}[t]{0.47\textwidth}
		\centering
		\scriptsize
		\includegraphics[width=1\textwidth]{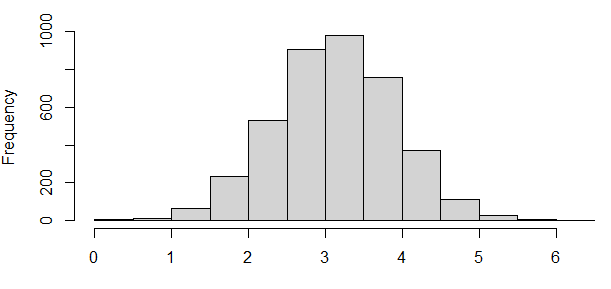}
		\caption{Beta Effort. \label{fig:beta_effort}}
	\end{subfigure}
	\label{histcont}
\end{figure}

Tables \ref{postmodelI} and \ref{postmodelII} show the results of Model I and II, respectively. The difference between them is the presence of $\beta_{atten}$ in the latter one.  The posteriors of model II, and the corresponding histograms, shown in Figures \ref{hist} and \ref{histcont}, indicate that $\beta_{prev}$ and $\beta_{effort}$ are the parameters distributed above zero. On the other hand, $\beta_{home}$ has a wider credible interval that includes zero value if $\beta_{atten}$ is included, as in the comparison between Model I and Model II. 

\begin{table}[ht]
	\centering
	\caption{Posterior Summary Statistics, Model I}
	\begin{tabular}{lrrrrrrr}
		\toprule
		Parameter & Rhat & n\_eff & mean & sd & 2.5\% & 50\% & 97.5\% \\ 
		\midrule
		b\_home & 1.000 & 7792 & 0.366 & 0.168 & 0.036 & 0.362 & 0.704 \\ 
		b\_prev & 1.001 & 3919 & 1.760 & 0.722 & 0.360 & 1.757 & 3.194 \\ 
		b\_effort & 1.000 & 7228 & 3.147 & 0.767 & 1.641 & 3.139 & 4.686 \\ 
		nu & 1.000 & 3048 & 12.375 & 5.036 & 5.057 & 11.566 & 24.360 \\ 
		sigma\_y & 1.000 & 4263 & 1.664 & 0.159 & 1.359 & 1.661 & 1.985 \\  
		\bottomrule
	\end{tabular}
	\label{postmodelI}
\end{table}

\begin{table}[ht]
	\centering
	\caption{Posterior Summary Statistics, Model II}
	\begin{tabular}{lrrrrrrr}
		\toprule
		Parameter & Rhat & n\_eff & mean & sd & 2.5\% & 50\% & 97.5\% \\ 
		\midrule
b\_home & 1.000 & 7025 & 0.324 & 0.175 & -0.028 & 0.327 & 0.657 \\ 
b\_prev & 1.000 & 4516 & 1.758 & 0.729 & 0.297 & 1.768 & 3.185 \\ 
b\_atten & 1.000 & 8830 & 0.376 & 0.496 & -0.588 & 0.370 & 1.354 \\ 
b\_effort & 1.000 & 9014 & 3.114 & 0.799 & 1.574 & 3.118 & 4.639 \\ 
nu & 1.000 & 3788 & 13.011 & 5.206 & 5.445 & 12.153 & 25.335 \\ 
sigma\_y & 1.000 & 5291 & 1.683 & 0.158 & 1.380 & 1.680 & 2.012 \\ 
		\bottomrule
	\end{tabular}
	\label{postmodelII}
\end{table}

Table \ref{postmodelIII} shows the difference of estimations when adding the $\beta_{day}$ parameter. The results seem to be stable, in the sense that the estimated values of $\beta_{prev}$ and $\beta_{effort}$ remain in similar intervals. We also show the result of changing the mean of the prior of $\nu$ to $1$.\\
 
Finally, Table \ref{postmodelIV} presents the results of Model IV. The ranking score is here based on the points (not the tries) of the previous season. We find in this case that the $\beta_{effort}$ parameter is similar as that found in the other models, while $\beta_{prev}$ has a lower mean.\\ 

\begin{table}[ht]
	\centering
	\caption{Posterior Summary Statistics, Model III}
	\begin{tabular}{lrrrrrrr}
		\toprule
		Parameter & Rhat & n\_eff & mean & sd & 2.5\% & 50\% & 97.5\% \\ 
		\midrule
		b\_home & 1.001 & 4146 & 0.322 & 0.320 & -0.302 & 0.321 & 0.944 \\ 
		b\_prev & 1.000 & 3782 & 1.721 & 0.727 & 0.295 & 1.717 & 3.147 \\ 
		b\_atten & 1.000 & 5355 & 0.279 & 0.480 & -0.670 & 0.276 & 1.249 \\ 
		b\_effort & 1.000 & 6561 & 3.064 & 0.770 & 1.571 & 3.066 & 4.538 \\ 
		b\_day & 1.001 & 3683 & -0.003 & 0.365 & -0.732 & -0.004 & 0.715 \\ 
		nu & 1.001 & 3424 & 6.876 & 2.301 & 3.355 & 6.498 & 12.389 \\ 
		sigma\_y & 1.000 & 3915 & 1.556 & 0.165 & 1.242 & 1.551 & 1.896 \\ 
		\bottomrule
	\end{tabular}
	\label{postmodelIII}
\end{table}

\begin{table}[ht]
	\centering
	\caption{Posterior Summary Statistics, Model IV}
	\begin{tabular}{lrrrrrrr}
		\toprule
		Parameter & Rhat & n\_eff & mean & sd & 2.5\% & 50\% & 97.5\% \\ 
		\midrule
		b\_home & 1.0 & 3328 & 0.3 & 0.3 & -0.3 & 0.3 & 0.9 \\ 
		b\_prev & 1.0 & 3071 & 1.1 & 0.7 & -0.3 & 1.1 & 2.5 \\ 
		b\_effort & 1.0 & 5604 & 3.1 & 0.8 & 1.6 & 3.1 & 4.6 \\ 
		b\_atten & 1.0 & 5391 & 0.2 & 0.5 & -0.7 & 0.2 & 1.2 \\ 
		b\_day & 1.0 & 3382 & 0.0 & 0.4 & -0.7 & 0.0 & 0.7 \\ 
		nu & 1.0 & 2681 & 5.2 & 1.4 & 2.9 & 5.1 & 8.4 \\ 
		sigma\_y & 1.0 & 2996 & 1.5 & 0.2 & 1.2 & 1.5 & 1.8 \\ 
		\bottomrule
	\end{tabular}
	\label{postmodelIV}
\end{table}

\begin{figure}[hbt!]
	\centering
	\caption{Bivariate relations between parameters}		
	\begin{subfigure}[t]{0.47\textwidth}
		\centering
		\scriptsize
		\includegraphics[width=1\textwidth]{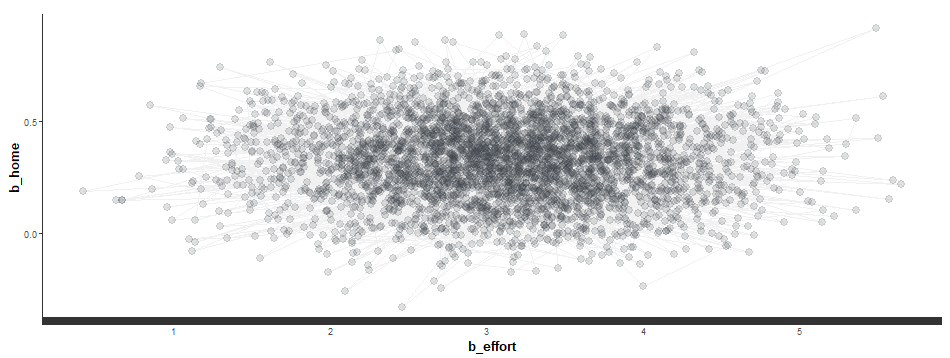}
		\caption{Beta Home versus Beta effort. \label{fig:effvshome}}
	\end{subfigure}%
	~ 
	\begin{subfigure}[t]{0.47\textwidth}
		\centering
		\scriptsize
		\includegraphics[width=1\textwidth]{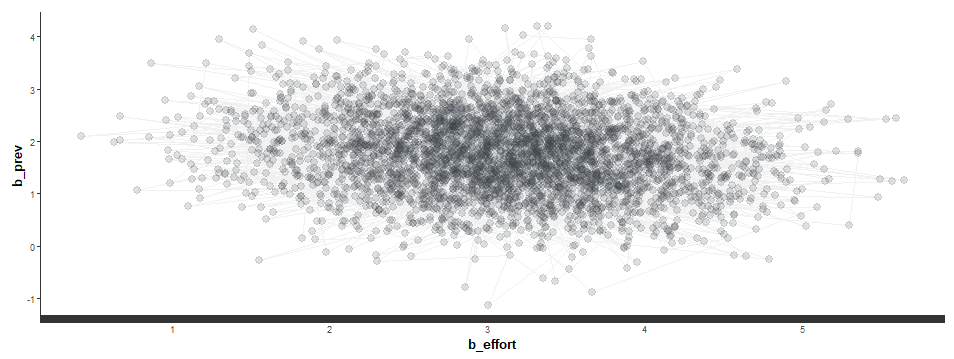}
		\caption{Beta Previous versus Beta effort. \label{fig:effvprev}}
	\end{subfigure}
	\label{biva}
\end{figure}

\begin{figure}[hbt!]
	\centering
	\caption{Simulated Score Difference Histogram}
	\includegraphics[width=0.75\textwidth]{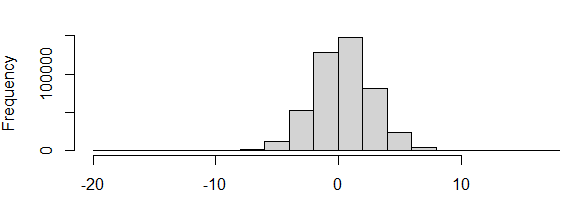}
	 \label{fig:score_diffsim}
\end{figure}

\begin{figure}[hbt!]
	\centering
	\caption{Histogram of replicated Score Differences}
	\includegraphics[width=1\textwidth]{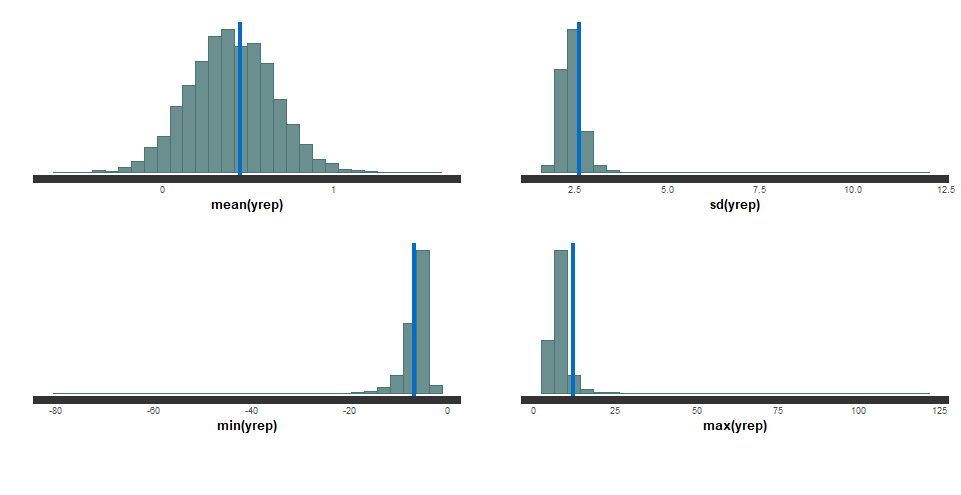}
	\label{fig:hist_yrep}
\end{figure}

\begin{figure}[hbt!]
	\centering
	\caption{Trace and Histogram of Nu}		
	\begin{subfigure}[t]{0.47\textwidth}
		\centering
		\scriptsize
		\includegraphics[width=1\textwidth]{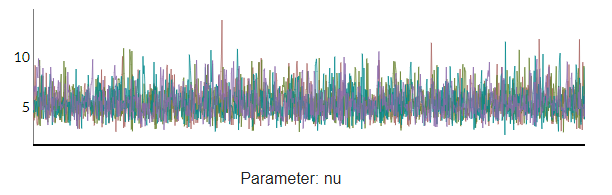}
		\caption{Trace. \label{fig:trace_nu}}
	\end{subfigure}%
	~ 
	\begin{subfigure}[t]{0.47\textwidth}
		\centering
		\scriptsize
		\includegraphics[width=1\textwidth]{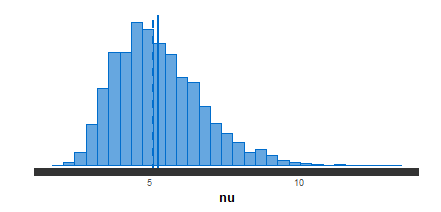}
		\caption{Histogram. \label{fig:hist_nu}}
	\end{subfigure}
	\label{tracenu}
\end{figure}

Besides the histograms corresponding to the values of the parameters , we can see in Figures \ref{biva} to \ref{tracenu}, further results of our model. Figure \ref{biva} indicates that there are no apparent relation between the values obtained for $\beta_{prev}$ and $\beta_{home}$ and those of $\beta_{effort}$. This is a strong suggestion that effort captures an effect that differs from both the ability of the team and the potential support (or antagonism) received in the field.\\
 
Figure \ref{fig:score_diffsim}, shows one of the 10000 simulated histograms of score differences. It can be compared to the original histogram corresponding to the actual championship. This is further detailed in Figure \ref{fig:hist_yrep}, where the histograms of the distributions of means and standard deviations, obtained in the replications, is depicted with gray bars, while the blue ones correspond to the values of the statistics computed from the observed data.\\

\begin{figure}[hbt!]
	\centering
	\caption{Trace and Histogram of Sigma}		
	\begin{subfigure}[t]{0.47\textwidth}
		\centering
		\scriptsize
		\includegraphics[width=1\textwidth]{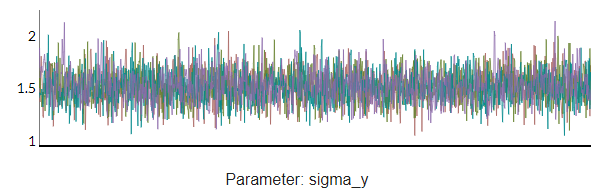}
		\caption{Trace. \label{fig:effvshome}}
	\end{subfigure}%
	~ 
	\begin{subfigure}[t]{0.47\textwidth}
		\centering
		\scriptsize
		\includegraphics[width=1\textwidth]{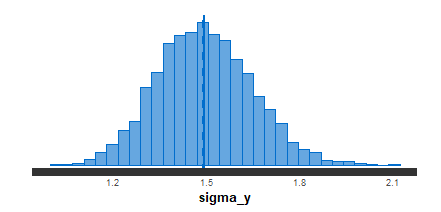}
		\caption{Histogram. \label{fig:effvprev}}
	\end{subfigure}
	\label{tracesigma}
\end{figure}

Notice that $\sigma$ and $\eta$ appear in the model only in a product and since both have Gaussian priors, it could be thought that they cannot be distinguished by likelihood. This would mean that only their product is identifiable but not the individual parameters. To check this we generate the trace-plots and histograms of these parameters. They are shown in Figures \ref{tracenu} and \ref{tracesigma}, indicating that this concern can be discarded since $\sigma$ and $\eta$ are identifiable.\\ 

Fioravanti et al. \cite{fioravanti2021home} claim that score differences in favor of the home team can vary when different parameters are taken into account. To explore this variability, we consider other prior distributions of $\beta_{home}$, with means $2$, $4$ and $6$. In all these cases, the results are similar to those obtained with mean $0.5$: the posterior converge to values close to $0.3$. \\


\section{Luck in games}
We assume that {\em luck} plays a significant role in games. According to the definition of Tango et al. \cite{tango2007book}, luck in a game can be understood as the difference between the actual performance observed and the ability of a team. In our case, we could identify it with the difference between the performance and both the effort and ability of a team. \\

For Tango and coauthors, the variability of luck is $\frac{p (1-p)}{g} $, where $ p = 0.5 $ and $ g = 22 $ is the number of games\footnote{The theoretical standard deviation of the distribution of luck over a season, using a binomial approximation to the normal distribution can be expressed as $\sqrt{\frac{p(1-p)}{g}}$, where $p = 0.500$ (since in average a team wins by chance half of the games), and $g =$ number of games.}. Then $Var(Luck) = 0.01136364 $. On the other hand, performance variability is the deviation in the number of games won by the teams in the league, yielding $ 0.03798835 $. Finally, effort variability can be captured by the variance of the effort ratio, $ 0.01563645 $. Then: 

$$Var(Performance) = Var(Luck) + Var(Effort) + Var(Ability)$$
$$ 0.03798835 = 0.01136364 + 0.01563645 + Var(Ability)$$

\noindent then, we find that:

$$ Var(Ability) = 0.01098826 $$

Then we can see that the variabilities in effort, ability and luck have slightly the same weight in the composition of the variability of performance.\\

An alternative definition of {\em luck} is that it arises when the residual in the regression of $y_g$ on the explanatory variables defined in Section 3 differs markedly from the mean value in the distribution of noises. That is, the presence of luck is revealed by a large impact of unobserved variables \cite{mauboussin2012success}. Elias et al. \cite{elias2012characteristics} and Gilbert and Wells \cite{gilbert2019ludometrics} consider four types of luck. The first one arises from physical randomization, by the use of dice, cards, etc. (I). The second kind of luck is due to simultaneous decision making (II). The third one is due to human performance fluctuating unpredictably (III), while the last one arises from matchmaking (IV).\footnote{A yellow card shown to a player (type II and III) can be seen as bad luck for that player's team and good luck for the rival. Although receiving a yellow or red card is somewhat under the control of a player, there are many situations where a lot of (bad) luck may be involved. Consider a player that it is in a perfect position for a tackle, but the player in front slips and end up receiving a hard hit in the head and get unconscious. While the slip mitigates the infraction, the tackler will be shown, at the very least, a yellow card. Having to play a game in the 15th round of the tournament between the top team and the bottom team means good luck for the top one and bad luck for the bottom team (Type IV).} \\


The underlying theoretical framework is very rich. In the sequel we give a simple example of the kinds of analysis that may be ultimately possible. To start, consider the types of luck that may have affected the outcomes:

\begin{itemize}
	\item Round 5 - London Wasps 34 vs 5 Exeter Chiefs: contrary to what happened in the previous week, the Wasps regained its captain and 3 players from the English national team, while Exeter missed 8 of its players who, after playing for the national team, were either injured or where forced to rest (type IV). Also, the Chiefs conceded 15 penalties (a high number for this level of competition) and were shown a yellow card, letting the Wasps score a try during the sin bin time (types II and III).\footnote{A yellow card shown to a player means that he must leave the field for  ten minutes (or is in the sin bin) while a red card means that he is expelled from the game.}
	\item Round 15 - Worcester Warriors 14 vs 62 Northampton Saints: the Warriors changed 10 players from the previous week's game, and had three injured players (the fullback and two tighthead props), while the Saints recovered two players from the national team. A victory would close the gap for the Saints on the top four (type II and IV). A Warriors player was shown a red card at the 49th minute, allowing 6 tries after that (type III).
	\item Round 20: Exeter Chiefs 74 vs 3 Newcastle Falcons: the Chiefs, already classified for the semifinals, needed a win to gain the home advantage during the playoffs. It was also the first game after more than 5 months that fans were allowed again to attend the play (type IV). Besides that, it was a record performance of Exeter, since it scored the largest difference in a top competition (type III). For the Falcons it was the longest trip of the tournament (type IV) and were shown a yellow card at minute 26, allowing two tries during the sin bin time (type II and III).
\end{itemize}

\section{Discussion}

In our analysis we found that while the results of rugby matches in the English Premiership can be explained by the ability of the teams, another highly significant variable is the motivation of players, reflected by the effort exerted by them. Luck, instead, seems to have had an impact only in games where the residuals are larger. \\

We have followed here Lenten and Winchester \cite{lenten2015secondary}, Butler et al. \cite{butler2020bonus} and Fioravanti et al. \cite{fioravanti2021effort}, assuming that the number of tries is an important component of any measure of effort. While the results obtained in our Bayesian analysis are sound, there exist many other ways of defining a proxy of motivation in Rugby. We can argue that an increased number of tackles indicates that a team has exerted a lot of effort, revealing that it is highly motivated. On the other hand, this team has not exerted a large effort in keeping the ball. One could also, with the help of GPS, track the {\em physical} effort of the players, and identify it with their motivation. In any case, the definition of a measure of effort as a proxy of motivation, like that of ability or luck has a degree of arbitrariness. \\

Future lines of research can be to explore and compare the impact of our proxy for motivation in other tournaments, and even compare the correct definition of  `effort' in different sports. Another topic that it is worth studying is the evolution of effort along time. The results of such investigation could be useful to assess how the incentives to the players may have changed, affecting the motivation of players. \\
\section*{Declarations}
\noindent\textbf{Funding} This research did not receive any specific grant from funding agencies in the public, commercial, or not-for-profit sectors.\\
\noindent\textbf{Conflicts of interest} The authors declare that they have no conflict of interest\\
\noindent\textbf{Acknowledgements} We would like to thank two anonymous reviewers and the Associate Editor for their comments, that were really helpful and constructive for the improvement of this work.
%

\bibliographystyle{abbrv}
\bibliography{ref}

\newpage

\section*{Appendix: STAN code}

\begin{lstlisting}
	data {
		int nteams; int ngames; int nweeks; 
		int home_week[ngames]; 	int away_week[ngames]; 
		int home_team[ngames]; 	int away_team[ngames]; 
		vector[ngames] score_diff;   row_vector[nteams] prev_perf; 
		vector[ngames] RatioH; 	vector[ngames] RatioA;
		vector[ngames] Att; vector[ngames] Day;  
	}
	parameters {
		real b_home; real b_prev; real b_effort; real b_att; real b_day; 
		real nu; real sigma_y; row_vector[nteams] sigma_a_raw; 
		matrix[nweeks,nteams] eta_a;        
	}
	transformed parameters {
		matrix[nweeks, nteams] a; 
		a[1,] = b_prev * prev_perf + eta_a[1,];
		for (w in 2:nweeks) {
			a[w,] = a[w-1,] + sigma_a_raw .* eta_a[w,];       
		}
	}
	model {
		vector[ngames] a_diff; vector[ngames] e_diff; 
		vector[ngames] home_adv; 
		nu ~ gamma(9,0.5); b_prev ~ normal(0.5,1);
		sigma_y ~ normal(0.5,1); b_home ~ normal(0.5,1);
		b_effort ~ normal(0.5,1); b_att ~ normal(0.5,1);
		b_day ~ normal(0.5,1);
		sigma_a_raw ~ normal(0,0.1); to_vector(eta_a) ~ normal(0,0.5);
		for (g in 1:ngames) {
			a_diff[g] = a[home_week[g],home_team[g]] - a[away_week[g],away_team[g]];
		}
		for (g in 1:ngames) {
			e_diff[g] = b_effort * (RatioH[g] - RatioA[g] );
		}
		for (g in 1:ngames) {
			home_adv[g] = b_home + b_att * Att[g] + b_day*Day[g];
		}
		score_diff ~ student_t(nu, a_diff + e_diff + home_adv, sigma_y);
	}
	generated quantities {
		vector[ngames] score_diff_rep;
		for (g in 1:ngames)
		score_diff_rep[g] = student_t_rng(nu, 
		(a[home_week[g] ,home_team[g]] - 
		a[away_week[g],away_team[g]]) 
		+ b_effort * (RatioH[g] - RatioA[g]) +
		 b_att * Att[g] + b_day*Day[g] + b_home, sigma_y);
	}
}
\end{lstlisting}

\end{document}